\documentclass[english,aps,preprint]{revtex4}
\usepackage[T1]{fontenc}
\usepackage[latin1]{inputenc}
\usepackage{amssymb}

\makeatletter

\usepackage{babel}
\makeatother
\begin{document}

\title{Classical and Quantum Analysis of One Dimensional Velocity Selection
for Ultracold Atoms}

\author{J.K. Fox, H.A. Kim$^{1}$, S.R. Mishra$^{2}$, S.H. Myrskog, A.M.
Jofre, L.R. Segal, J.B. Kim$^{1}$, and A.M. Steinberg}

\affiliation{Department of Physics, University of Toronto\\
 Toronto, Ontario M5S 1A7 Canada\\
 $^{1}$Permanent address: Department of Physics, Korea National University
of \\
 Education,\\
 Chungbuk, Korea 363-791\\
 $^{2}$Permanent address: Center for Advanced Technology, Indore
452013, India }

\date{February 21, 2004}


\begin{abstract}
We discuss a velocity selection technique for obtaining cold atoms,
in which all atoms below a certain energy are spatially selected from
the surrounding atom cloud. \ Velocity selection can in some cases
be more efficient than other cooling techniques for the preparation
of ultracold atom clouds in one dimension. \ With quantum mechanical
and classical simulations and theory we present a scheme using a dipole
force barrier to select the coldest atoms from a magnetically trapped
atom cloud. \ The dipole and magnetic potentials create a local minimum
which traps the coldest atoms. \ A unique advantage of this technique
is the sharp cut-off in the velocity distribution of the sample of
selected atoms. \ Such a non-thermal distribution should prove useful
for a variety of experiments, including proposed studies of atomic
tunneling and scattering from quantum potentials. \ We show that
when the $rms$ size of the atom cloud is smaller than the local minimum
in which the selected atoms are trapped, the velocity selection technique
can be more efficient in 1-D than some common techniques such as evaporative
cooling. For example, one simulation shows nearly $6\%$ of the atoms
retained at a temperature $100$ times lower than the starting condition. 
\end{abstract}
\maketitle

\section{Introduction}

Advances in laser cooling of neutral atoms during the last two decades
have led to observations of many new phenomena\cite{cornell bec,ketterle,bloch,jin}
as well as the development of a variety of atom-optics technologies\cite{phillips 4wm,esslinger,phillips al,dalibard,hinds mir,mlynek bs,prentiss,ertmer}.
\ To study some of these effects, such as quantum tunneling, and
studies in quantum chaos, only require that the atoms being studied
be cold in one dimension\cite{ae super,ae annalen,raizen tun,raizen chaos}.
\ For such studies, cooling atoms in only one dimension should serve
the purpose and can prove more efficient than cooling in all three
dimensions. \ Cooling in one dimension has been previously studied
in processes such as `delta kick' cooling, which compresses the velocity
distribution of an atom cloud with magnetic field gradients, paid
for with an increase in the spatial distribution, thus conserving
phase space density\cite{stef}. \ As well, atoms have been cooled
to the ground state of a one dimensional optical lattice\cite{jessen,weiss}.

In this paper, we analyze an efficient and straightforward method
to achieve low one dimensional temperatures in neutral atoms through
velocity selection. \ The technique takes advantage of the fact that
in an atom cloud with a Gaussian velocity distribution there is a
large population near zero velocity along any individual direction.
\ By utilizing the dipole force to manipulate the motion and spatial
position of atoms, a process dependent on the atoms' own energies,
cold atoms can be separated from their high energy counterparts. \ Manipulation
of atoms with the dipole force is a common experimental technique
and has been used in several ways such as atom trapping\cite{grimm,ozeri},
reflection of atoms\cite{dalibard}, quantum state engineering\cite{salomon},
and a similar selection of low energy atoms in an optical lattice\cite{raizen v,mylnek}.
\ We show here that this version of velocity selection can not only
provide cold samples of atoms at high efficiencies, but even pure
1-D quantum states at high efficiencies when in the ultracold regime
of $\sim10nK$. Experimental studies on this technique were performed
in this laboratory in parallel with this theoretical study\cite{vsel}.

\section{Velocity Selection Overview}

The proposed velocity selection process is the following: atoms are
trapped in a weak quadrupole magnetic trap, while a blue-detuned laser
sheet sweeps through the trap. \ The repulsive potential due to the
blue-detuned laser pushes the colder atoms (those with less kinetic
energy than the laser potential) away from the more energetic atoms
which can classically overcome this potential. \ The result is a
cold sample of atoms spatially separated from the original atom cloud.
\ Away from the center of the magnetic trap the magnetic potential
is approximately separable allowing us to express the combined magnetic
and optical potential in one dimension. \ (This approximation could
be made exact by using a harmonic rather than quadrupole potential.)
\ A schematic of this is shown in figure 1. \ The combination of
the magnetic and dipole potentials results in a potential of \begin{equation}
U=U_{M}+U_{D}=\mu B^{\prime}\left|x\right|+U_{0}\exp\left[-\frac{(x-x_{D})^{2}}{2w_{0}^{2}}\right]\label{1}\end{equation}
 where $\mu$ is the atomic magnetic moment, $B^{\prime}$ is the
spatial gradient of the magnetic trap, $x_{D}$ is the position of
the dipole barrier, $w_{0}$ is the $rms$ width of the dipole barrier,
and $U_{0}$ is the peak of the dipole force potential. \ To make
the above potential conservative, the detuning of the laser from the
atom's resonant frequency must be large enough so that photon absorption
by the atoms is negligible. \ In typical dipole experiments in which
conservative potentials are required, a detuning of $\delta\gtrsim1000\Gamma$
is sufficient, where $\Gamma$ is the natural linewidth of the atom.
\ This combination of the magnetic and dipole potentials is shown
in figure 2. \ Note from figure 2b that the well depth is not equal
to the height of the dipole potential barrier, $U_{0}$, but rather
$U_{eff}$, which is somewhat smaller due to the finite width of the
barrier.

After the selection process shown in figure 1, the selected atoms
find themselves in the potential well bounded on one side by the magnetic
trap and on the other by the dipole force beam. \ The atomic velocity
distribution of this selected cloud is a truncated Gaussian, with
no atoms having energy greater than the depth of the potential well.
\
It's important to note that when an atom cloud is selected at a temperature
we define as $T_{eff}\equiv2U_{eff}/k_{B}$, where $k_{B}$ is Boltzmann's
constant, this does not describe the $rms$ kinetic energy, but the
absolute maximum kinetic energy of the hottest atoms in the sample.
\ Such a truncated distribution is required, in particular, for a
planned experiment studying the tunneling of $^{85}Rb$ through a
relatively macroscopic barrier\cite{ae annalen}. \ It must be known
for certain that no atoms in the sample have energy greater than the
barrier height, or else they could traverse the barrier classically.
\ In a thermal distribution, there would always be the possibility
of classical traversal by atoms in the high energy tails of the distribution.
Similar considerations apply to other experiments, such as proposed
studies of quantum potential scattering\cite{muga}.

\section{Classical Theory and Simulation}

Until the temperature of the atom cloud is in the ultra low regime
where quantum effects begin to dominate, the velocity selection process
can be largely described by classical physics. \ By considering the
magnetic and dipole potentials as classical, conservative potentials,
the dependence of the efficiency of the selection process on various
parameters can be studied when still in this regime.

We begin with a description of the `experiment' and analytic estimates
of selection efficiency before presenting the results of the classical
simulations. \ We start with an atom cloud with Gaussian spatial
and velocity distributions of $rms$ radius $r_{0}$ and $rms$ velocity
$v_{0}$ respectively. \ The initial temperature is $T_{0}=mv_{0}^{2}/k_{B}$.

We model a dipole force beam moving slowly through the atom cloud.
\ When the dipole beam is moved slowly enough, whether or not an
individual atom is swept up by the barrier depends only on whether
its total energy is greater than that of the potential created by
the barrier. \ Of principal interest is how the efficiency of the
selection process depends on barrier height for different atom cloud
sizes, temperatures, and magnetic field gradients. \ Three separate
regimes occur, in which the dependence of efficiency on barrier height
differs drastically. \ Which regime the experiment is in depends
on the ratio of the initial kinetic energy $(\left\langle KE\right\rangle _{i}=\frac{1}{2}mv_{0}^{2})$
to the variation of magnetic potential energy across the cloud $(\left\langle PE\right\rangle _{i}=\sqrt{2/\pi}\mu B^{\prime}r_{0})$.

The first regime $(\left\langle KE\right\rangle _{i}\gg\left\langle PE\right\rangle _{i})$
occurs when the $rms$ radius of the atom cloud is much smaller than
the size of the potential well formed by the magnetic gradient and
dipole barrier $(r_{0}\ll r_{w})$. Since the magnetic trap is linear
in position, $r_{w}\sim U_{eff}/(\mu B^{\prime})$. \ When $r_{0}\ll r_{w}$
very few atoms are given significant potential energy by the magnetic
gradient, as most atoms are always near the minimum of the potential
well. \ Therefore the velocity selection process selects atoms by
their kinetic energy only. \ The requirement for selection of an
atom in this regime is merely $\frac{1}{2}mv^{2}<U_{eff}$, or $v<v_{c}\equiv\sqrt{2U_{eff}/m}=\sqrt{k_{B}T_{eff}/m}.$
We define the efficiency of the process to be the ratio of number
of atoms selected during the process to the total number of atoms
in the cloud. In this regime this efficiency is given by an error
function of the original Gaussian velocity distribution \begin{equation}
\eta_{KE}=\frac{1}{\sqrt{2\pi}v_{0}}\int_{-v_{c}}^{+v_{c}}\exp\left[-\frac{v^{2}}{2v_{0}^{2}}\right]dv\label{2}\end{equation}
 At low barrier heights, that is, for $T_{eff}<T_{0}$, this can be
approximated by \begin{equation}
\eta_{KE}\simeq\sqrt{\frac{2}{\pi}}\left(\frac{v_{c}}{v_{0}}\right)=\sqrt{\frac{2T_{eff}}{\pi T_{0}}}.\label{3}\end{equation}
 This is the desired regime for the experiment, as the efficiency
of the selection process may be high even for large temperature ratios.
\ For example, atoms as cold as $10nK$ (similar to temperatures
required for future experiments) could be obtained from an atom cloud
at $1\mu K$ with an efficiency of $\sim8\%$. \ This is to be contrasted
with typical efficiencies of $<1\%$ for evaporative cooling by a
factor of $100$ in temperature\cite{cornell,ketterle 2}

The other extreme occurs when $r_{0}\gg r_{w}$ $(\left\langle KE\right\rangle _{i}\lesssim\left\langle PE\right\rangle _{i})$,
that is, when the atom cloud size is much greater than the size of
the potential well. \ In this case, even though the initial atom
temperature, $T_{0}$, may be much less than the dipole potential
height, a significant amount of potential energy is given to many
of the atoms by the magnetic gradient. \ The requirement for selection
of an atom now involves both the kinetic and potential energy of the
atom, $U_{a}+KE_{a}<U_{eff}$. \ During the selection process, atoms
are being selected by their kinetic energy as well as their potential
energy, given by their position in the potential well. \
Consider the requirement for an atom of zero kinetic energy to be
selected. \ The atom must be close enough to the well minimum that
its energy is less than the well depth; $\mu_{B}B^{\prime}x<U_{eff}$,
or $x<x_{c}\simeq U_{eff}/(\mu_{B}B^{\prime})=k_{B}T_{eff}/(2\mu_{B}B^{\prime})$.
\ For an atom cloud with a gaussian spatial distribution then, the
efficiency of selection, in terms of the spatial component only, is
another error function integral: \begin{equation}
\eta_{PE}=\frac{1}{\sqrt{2\pi}r_{0}}\int_{x_{D}}^{x_{c}}\exp\left[-\frac{x^{2}}{2r_{0}^{2}}\right]dx\label{4}\end{equation}
 Again, for low barrier heights such that $T_{eff}<T_{0}$, this can
be approximated as $\eta_{PE}\simeq\sqrt{2/\pi}(x_{c}/r_{0})\simeq\sqrt{2/\pi}T_{eff}k_{B}/(2r_{0}\mu_{B}B^{\prime})$.
\ In conjunction with the square root dependence of efficiency on
barrier height when only the kinetic energy is being considered, the
expected efficiency, in the low barrier limit, becomes $\eta\simeq\beta T_{eff}^{3/2}$,
where $\beta=k_{B}/(T_{0}^{1/2}r_{0}\pi\mu_{B}B^{\prime})$. \ A
dependence of this sort is less desirable than that of equation (3),
as the efficiency falls off faster than the final temperature. \ (Note
that in evaporative cooling, efficiency is typically roughly linear
in temperature.)

In between these two regimes, the efficiency should be approximately
linear with barrier height, $\eta\propto T_{eff}$. \ The exact expression
for the efficiency in all cases for this description of the selection
process is\begin{equation}
\eta=\frac{1}{2\pi v_{0}r_{0}}\int_{-v_{c}}^{v_{c}}\exp\left[-\frac{v^{2}}{2v_{0}^{2}}\right]\int_{x_{D}}^{x=x_{c}}\exp\left[-\frac{x^{2}}{2r_{0}^{2}}\right]dxdv.\label{5}\end{equation}
 Note that since, in the general case, the critical position for selection
is $x_{c}=(U_{eff}-\frac{1}{2}mv^{2})/(\mu_{B}B^{\prime})$, this
equation is quite convoluted. \ This expression has been numerically
evaluated for various regimes and gives the same qualitative dependences
as the actual classical simulation.

Figure 3 shows the results of the simulation for selection efficiency
versus barrier height for a very small atom cloud $(r_{0}=25\mu m)$
of temperature $T_{0}=50\mu K$ in a well made of a magnetic gradient
of $B^{\prime}=0.5G/cm$ , and a dipole barrier with an $rms$ radius
of $5\mu m$. For comparison, the solid line presents the results
of equation (3). For these parameters $\left\langle KE_{i}\right\rangle =373\left\langle PE_{i}\right\rangle $.
\ The graph definitely shows an $\eta\sim\sqrt{T_{eff}/T_{0}}$ dependence
of efficiency on barrier height. \ As many as $5\%$ of the atoms
may be selected at temperatures as low as $T_{eff}=500nK=T_{0}/100$.

Next, shown in figure 4, is the efficiency versus barrier height for
a much larger magnetic gradient of $100G/cm$, with the other parameters
unchanged. \ Now $\left\langle KE_{i}\right\rangle =2\left\langle PE_{i}\right\rangle $.
This shows much lower efficiency because of the spatial dependence
now present. The accompanying solid line shows the $\beta T_{eff}^{3/2}$
dependence suitable for this regime. Note from the condition for the
large cloud size limit, $r_{0}\gg U_{eff}/(\mu_{B}B^{\prime})$, that
a larger initial atom cloud size, as opposed to a larger magnetic
field, would also create an efficiency of this dependence.

\section{Quantum Theory and Simulations}

It is at ultra low temperatures that quantum effects start to become
evident. \ Performing simulations using quantum theory allows us
to explore these low energy effects not predicted by the classical
description. \
Instead of considering a classical Gaussian cloud of atoms, we now
start with an initial wavefunction $\psi(x,t=0)$ given by \begin{equation}
\psi(x,t=0)=\sum_{n}\psi_{n}(x,t=0)P_{n}^{1/2}e^{i\phi_{n}}=\sum_{n}\left(\frac{1}{2\pi x_{0}^{2}}\right)^{1/4}\exp\left[-\frac{\left(x-x_{n}\right)^{2}}{4x_{0}^{2}}\right]P_{n}^{1/2}(x_{n})e^{i\phi_{n}}\label{6}\end{equation}
 where $P_{n}(x_{n})=\left(\frac{1}{2\pi r_{0}^{2}}\right)^{1/2}\exp\left[-\frac{x_{n}^{2}}{2r_{0}^{2}}\right]$
is the weighting due to the overall gaussian distribution of the ensemble,
$\psi_{n}(x,t=0)=\left(\frac{1}{2\pi x_{0}^{2}}\right)^{1/4}\exp\left[-\frac{\left(x-x_{n}\right)^{2}}{4x_{0}^{2}}\right]$,
and $\phi_{n}$ is a random phase. \ Random phases were given to
each component to ensure that there is no phase coherence between
individual components, making the initial wavefunction a collection
of highly localized energetic atoms. \
To make sure this is the case we analyze the initial momentum distribution
with a Fourier transform and check that it resembles a fairly smooth
gaussian distribution of $rms$ momentum width $p_{0}=\hbar/(2x_{0})$.
\ The result is one representation of a thermal cloud of atoms of
$rms$ radius $r_{0}$, and initial kinetic energy $\left\langle KE\right\rangle _{i}=p_{0}^{2}/(2m)$.
For these simulations we use the mass of $^{85}Rb$. The wavefunction
$\psi(x,t)$ obeys the time dependent Schrodinger equation $i\hbar\frac{\partial}{\partial t}\psi(x,t)=\left(H_{0}+U(x,t)\right)\psi(x,t)$
where $H_{0}=-\hbar^{2}/(2m)\partial^{2}/\partial x^{2}$ and $U(x,t)$
is the combined magnetic and dipole potential. \ The solution to
the time dependent Schrodinger equation is found by using the split-operator
method and evolving the initial wavefunction in time by \[
\psi(x,t+\Delta t)=\exp\left[-\frac{i}{\hbar}\int_{t}^{t+\Delta t}\left(H_{0}+U(x,t)\right)dt\right]\psi(x,t)\]

\begin{equation}
=\exp\left[-\frac{i\hbar}{4m}\Delta t\frac{\partial^{2}}{\partial x^{2}}\right]\exp\left[-\frac{i}{\hbar}\Delta tU(x,t)\right]\exp\left[-\frac{i\hbar}{4m}\Delta t\frac{\partial^{2}}{\partial x^{2}}\right]\psi(x,t).\label{7}\end{equation}
 This equation is evaluated in momentum space. \ A Fourier transform
then yields the final real space wavefunction $\psi(x,t_{f})$. \ The
final result of the velocity selection is then found from $\left|\psi(x,t_{f})\right|^{2}$
to obtain the probability distribution. Because of the random phases
between each $\psi_{n}(x,t_{f})e^{i\phi_{n}}$, $\left|\psi(x,t_{f})\right|^{2}=\left|\sum_{n}P_{n}^{1/2}(x_{n})\psi_{n}(x,t_{f})e^{i\phi_{n}}\right|^{2}$
can also be written as $\sum_{n}P_{n}(x_{n})\left|\psi_{n}(x,t_{f})\right|^{2}$,
making clear the connection of this approach to a more computationally
intense density-matrix approach.

Shown in figure 5 is a plot of the total probability distribution
$\left|\Psi(x,t_{f})\right|^{2}$ after selection in a magnetic trap
of $10nK/\mu m$ gradient along with a dipole barrier with a $97nK$
depth and $5\mu m$ $rms$ width which, together, make a potential
well of $4.6nK$. Also shown is the potential, $U(x,t_{f})$, showing
a local minimum at $27\mu m$ in which selected atoms are trapped.
\ The original atom cloud has an $rms$ radius of $r_{0}=8\mu m$,
while the $rms$ width of each individual wavepacket, $\psi_{n}(x,t=0)$,
is $x_{0}=0.07\mu m$, corresponding to a temperature of $T_{0}=292nK$
(a reasonable starting point after techniques such as delta-kick cooling)\cite{stef}.
\ The result of the selection is a thermal atom cloud which has expanded
in the weak magnetic field due to its kinetic energy. At the local
minimum is the selected portion of the wavefunction, representing
$5.8\%$ of the original wavefunction and having a width of $\sim2\mu m$.
\ The energy of the selected atom cloud, $\psi_{s}(x)$, is found
by projecting out $\psi_{s}(x)$ and calculating the expectation values
of its potential and kinetic energy. \ Potential energy is given
by \begin{equation}
\left\langle PE_{s}\right\rangle =\int\left|\psi_{s}(x)\right|^{2}\left[U(x,t_{f})-U_{min}\right]dx\label{9}\end{equation}
 where the low point of the well $U_{min}$ is subtracted so that
the result is with respect to the local minimum. \ For the kinetic
energy, a transform into momentum space, $\tilde{\psi_{s}}(p)$, is
made and the expression \begin{equation}
\left\langle KE_{s}\right\rangle =\int\left|\tilde{\psi_{s}}(p)\right|^{2}\frac{p^{2}}{2m}dp\label{10}\end{equation}
 is evaluated. \ The results for the wavepacket shown in figure 5
are a potential energy of $1.6nK$ and a kinetic energy of $1.15nK$.
\ Close to the minimum of the potential well, the shape is very close
to a harmonic potential. \ If a harmonic approximation is made, one
obtains a trap frequency (for $^{85}Rb$) of $280Hz$ corresponding
to a ground state energy of $2.1nK$, and first excited state energy
of. $3.15nK$\ \ \ While the selected wavefunction shown in figure
5 certainly looks like the ground state of a harmonic oscillator,
the difference in potential and kinetic energy in the cloud and its
difference in total energy from the expected ground state energy points
to some slight anharmonicity in the potential well as well as a slight
mixture with the first excited state.

Shown in figure 6 is a similar plot with the same atomic-cloud parameters
as for figure 5, but here a larger dipole potential of $107nK$ creates
a deeper well of $U_{eff}=9.8nK$. \ The result is a slightly larger
selected wavefunction no longer resembling the ground state of the
well.

The quantum simulation also yields a prediction for the expected selection
efficiencies. \ By integrating over the probability distribution
for the selected cloud, $\left|\psi_{s}(x)\right|^{2}$, the efficiency
of selection for the $4.6nK$ and $9.8nK$ well depths shown in figures
5 and 6 were found to be $5.8\%$ and $8.5\%$ respectively. \ By
varying the height of the dipole barrier, a graph of the efficiency
versus well depth may be constructed with the quantum theory for comparison
to the classical predictions. \ Figure 7 shows the dependence of
the efficiency on the well depth, $U_{eff}$ for the same parameters
as used for figures 5 and 6 but with a varying potential. \ Also
shown, in dashed lines, is the efficiency versus well depth as calculated
by the classical simulation for the same parameters. \ In the inset
are shown the results of the quantum simulation for very small well
depths. The quantum simulation shows general agreement with the classical
simulations. For low dipole barriers however, the quantum simulation
produces an efficiency curve with some step structure, indicating
energy gaps for which an increase in barrier height does not show
a corresponding efficiency increase.

\section{Conclusion}

We have proposed a method of not only achieving ultra-low one dimensional
temperatures in samples of atoms, but also of creating specific quantum
states and nonthermal mixtures in a potential well. \
For very small clouds of $\lesssim50\mu m$ a large number ($\sim5\%$)
of atoms can be selected with an efficiency which varies as the square
root of the final temperature. \ This is much more economical than
techniques whose efficiency drops linearly with temperature. \ The
technique also guarantees a maximum energy set by the parameters of
the experiment. \ This sharp cut-off in the velocity distribution
is useful when performing experiments in which the energy of each
atom must be below a certain limit, where a Boltzmann distribution
would be unsatisfactory. \ Our theory and simulations show that there
are experimentally accessible parameter regimes where velocity selection
can be more efficient in terms of final atom number than existing
cooling methods at similar temperatures. Atoms at temperatures of
$\lesssim20nK$ are shown to be in non-classical states, but as higher
temperatures are selected, the resultant cloud is increasingly classical.
These characteristics of the velocity selection technique provide
the motivation for present experimental work on the method\cite{vsel}.

\section{Acknowledgements}

We would like to thank the Natural Sciences Engineering and Research
Council of Canada, the Canadian Foundation for Innovation, the Canadian
Institute for Photonics Innovation and the Ontario Research and Development
Challenge Fund. We also thank Paul Brumer for the use of his computational
facilities.

 \newpage

Figure 1: \ Schematic of velocity selection process. \ Atoms are
first placed in a shallow magnetic trap. \ Moving dipole potential
then sweeps cold atoms up the magnetic potential away from more energetic
atoms.

Figure 2: \ Shape of combined potential of quadrupole magnetic fields
and blue detuned dipole force laser beam. \ Close-up shows effect
of gradient of magnetic potential and width of dipole beam on the
well delpth.

Figure 3: \ Simulated efficiency versus barrier height for a $25\mu m$,
$50\mu K$ atom cloud in a $0.5G/cm$ magnetic potential. For low
$U_{eff}$ ($\ll k_{B}T_{0}$), the efficiency can be approximated
by the analytic expression $\eta=\sqrt{2T_{eff}/(\pi T_{0})}$, shown
by the solid line. This favorable dependence is seen for parameters
such that $\mu_{B}B^{\prime}r_{0}\ll U_{eff}$, illustrating the regime
where negligible potential energy is given to the atom cloud by the
magnetic trap potential.

Figure 4: \ Simulated efficiency dependence graph for a $25\mu m$,
$50\mu K$ atom cloud in a $100G/cm$ magnetic potential. \ With
these parameters, $\mu_{B}B^{\prime}r_{0}\gg U_{eff}$ and efficiency
now has a dependence on both the velocity and spatial components of
the atom cloud, resulting in a dependence at low well depths which
can be approximated by $\eta=\beta T_{eff}{}^{3/2}$ which is shown
as the solid line for comparison.

Figure 5. \ Probability distribution, $\left|\Psi(x,t_{f})\right|^{2}$,
of a thermal atomic cloud after velocity selection. \ The selected
cloud, $\psi_{s}(x)$, is towards the left of the plot in a $4.6nK$
local minimum, spatially separated from the rest of the more energetic
atoms remaining in the center of the trap. \ $5.8\%$ of the original
atom cloud was selected to form this wavepacket having potential and
kinetic energies of $1.6nK$ and $1.15nK$ respectively. \ The initial
atom cloud had a $rms$ size of $8\mu m$ and an $rms$ temperature
of $292nK$ making a ratio of final to initial temperature $T_{f}/T_{0}\simeq1/100$.
\ Note the resemblence of the selected probability distribution to
that of a single pure quantum state in this harmonic-like potential.
\ Also shown is the combination of the magnetic and dipole potentials
showing a shallow well where the selected wavefunction is resting.

Figure 6: \ Plot of $\left|\Psi(x,t_{f})\right|^{2}$ for a deeper
potential well of $9.8nK$. \ Atoms were selected into the local
minimum here with an efficiency of $8.5\%$ and a ratio of final to
initial temperature of $T_{f}/T_{0}\simeq1/30$. \ At this deeper
selection potential more than one bound state is observed as shown
by the structure in the selected wavepacket.

Figure 7: \ A plot of efficiency of selection versus the potential
depth, $U_{eff}$ calculated from the quantum simulation (solid line)
as well as from the classical simulation (dashed line) for comparison.
Insets show the selected wavefunctions, $\psi_{s}(x)$, from figures
5 and 6 and their relative positions on the efficiency curve. 
\end{document}